\documentclass[preprint,aps,nofootinbib,showpacs,amsfonts,psfig]{revtex4}

\input{epsf.tex}

\newcommand{\be}{\begin{equation}}
\newcommand{\ee}{\end{equation}}
\newcommand{\bea}{\begin{eqnarray}}
\newcommand{\eea}{\end{eqnarray}}
\newcommand{\ba}{\begin{array}}
\newcommand{\ea}{\end{array}}

\begin{document}
\draft
\title{Kinematic Constraints to the Transition Redshift from SNe Ia Union Data}

\author{J. V. Cunha$\,^1$}
\address{$^1$Departamento de Astronomia, Universidade de S\~ao Paulo, USP,
05508-900 S\~ao Paulo, SP, Brazil}

\begin{abstract}
The kinematic approach to cosmological tests provides a direct
evidence to the present accelerating stage of the universe which
does not depend on the validity of general relativity, as well as on
the matter-energy content of the Universe. In this context, we
consider here a linear two-parameter expansion for the decelerating
parameter, $q(z)=q_0+q_1z$, where $q_0$ and $q_1$ are arbitrary
constants to be constrained by the Union supernovae data. By
assuming a flat Universe we find that the best fit to the pair of
free parameters is ($q_0,q_1$) = ($-0.73,1.5)$ whereas the
transition redshift is $z_t = 0.49^{+0.14}_{-0.07}$ ($1\sigma$)
$^{+0.54}_{-0.12}$ ($2\sigma$). This kinematic result is in
agreement with some independent analyzes and accommodates more
easily many dynamical flat models (like $\Lambda$CDM).
\end{abstract}


\maketitle

%\twocolumn

\section{Introduction}

It is now widely believed that the Universe at redshifts smaller
than unity underwent a ``dynamic phase transition" from decelerating
to accelerating expansion which has been corroborated by several
independent analyzes. In the context of the general relativity
theory such a phenomenon can be interpreted as a dynamic influence
of some sort of dark energy whose main effect is to change the sign
of the universal decelerating parameter $q(z)$.

The most direct observation supporting the present accelerating
stage of the Universe comes from the luminosity distance versus
redshift relation measurements using supernovas (SNe) type Ia
\cite{Riess,Kowalski08}. Initially, they were interpreted in light
of $\Lambda$CDM scenarios using either background or inhomogeneous
luminosity distances \cite{SantosLimaCunha08}. However, independent
theoretical and observational/statistical analyses point to more
general models whose basic ingredient is a negative-pressure dark
energy component \cite{Padm03}.

Nowadays, the most accepted cosmic picture is an expanding flat (or
nearly flat) spatial geometry whose dynamics is driven by an exotic
component called dark energy, 3/4 of composition, and 1/4 for matter
component (baryons plus dark). Among a number of possibilities to
describe this dark energy component, the simplest and most
theoretically appealing way is by means of a positive cosmological
constant $\Lambda$. Other possible candidates are:  a vacuum
decaying energy density, or a time varying $\Lambda$-term
\cite{Lambdat}, a time varying relic scalar field slowly rolling
down its potential \cite{CampEscal}, the so-called ``X-matter", an
extra component simply characterized by an equation of state $p_{\rm
x}=\omega\rho_{\rm x}$ \cite{xmatt}, the Chaplygin gas whose
equation of state is given by $p= -A/\rho$ where $A$ is a positive
constant \cite{chaplygin}. For scalar field and XCDM scenarios, the
$\omega$ parameter may be a function of the redshift
\cite{Efstat99Cunha07TurRie02}, or still, as it has been recently
discussed, it may violate the dominant energy condition and assume
values $<-1$ when the extra component is named phantom cosmology
\cite{Cald02Lima03}. It should be stressed, however, that all these
models are based on the validity of general relativity or some of
its scalar-tensorial generalizations.

On the other hand, Turner and Riess \cite{TurRie02} have discussed
an alternative route - sometimes called kinematic approach - in
order to obtain information about the beginning of the present
accelerating stage of the Universe with no assumption concerning the
validity of general relativity or even of any particular metric
gravitational theory (in this connection see also Weinberg
\cite{Weinb72}). Although considering that such a method does not
shed light on the physical or geometrical properties of the new
energetic component causing the acceleration, it allows one to
assess the direct empirical evidence for the transition
deceleration/acceleration in the past, as provided by SNe type Ia
measurements. Many authors have constrained values for the
transition redshift ($z_t$), explored implications on the cosmic
acceleration, or yet, used it as trustworthy discriminator for
cosmology. This value is obtained without supposing any energy
components (baryons, dark matter, dark energy), or any other cause
for acceleration
\cite{KinemWork,KinemWork2,KinemWork3,KinemWork4,CunLim08}.

More recently, Mortsell and Clarkson \cite{Clarkson08} applied a
kinematic approach to determine if the Copernican assumption is
violated. Moreover, by using a Taylor expansion of the scale factor,
it was found  that the acceleration today is detected to an accuracy
$> 12\sigma$. It was also claimed with basis  on the ratio of  the
scale of the baryon acoustic oscillations as imprinted in the cosmic
microwave background and in the large scale distribution of
galaxies, that a flat or negatively curved universe decelerate at
high redshifts.

In this paper, by adopting  the kinematic approach for which the
full gravitational theory also does not play  a prominent role, we
investigate the cosmological implications on the transition redshift
$z_t$ and deceleration parameters from the Supernovae Cosmology
Project (SCP) Union sample \cite{Kowalski08}.

\section{Luminosity distance and sample}

To begin with, let us  assume that the spatially flat
Friedman-Robertson-Walker (FRW) metric geometry, as motivated by
inflation and the WMAP results \cite{Komat08}. Following standard
lines \cite{TurRie02}, the luminosity distance is kinematically
defined by the following integral expression (in our units  $c=1$).
\begin{eqnarray}\label{eq:dLq}
D_L(z) = (1+z)\int_0^z {du\over H(u)} = \frac{(1+z)}{H_0} \hspace{1.5cm}& \nonumber \\
\int_0^z \exp{\left[-\int_0^u [1+q(u)]d\ln (1+u)\right]} du, &
\end{eqnarray}
where the $H(z)=\dot a/a$ is the Hubble parameter, and, $q(z)$, the
deceleration parameter, is defined by
\begin{eqnarray}\label{qz}
q(z)\equiv -\frac{a\ddot a}{\dot a^2} = \frac{d H^{-1}(z)}{ dt} -1.
\end{eqnarray}

\begin{figure}[htb]
\centerline{\epsfysize=90mm\epsffile{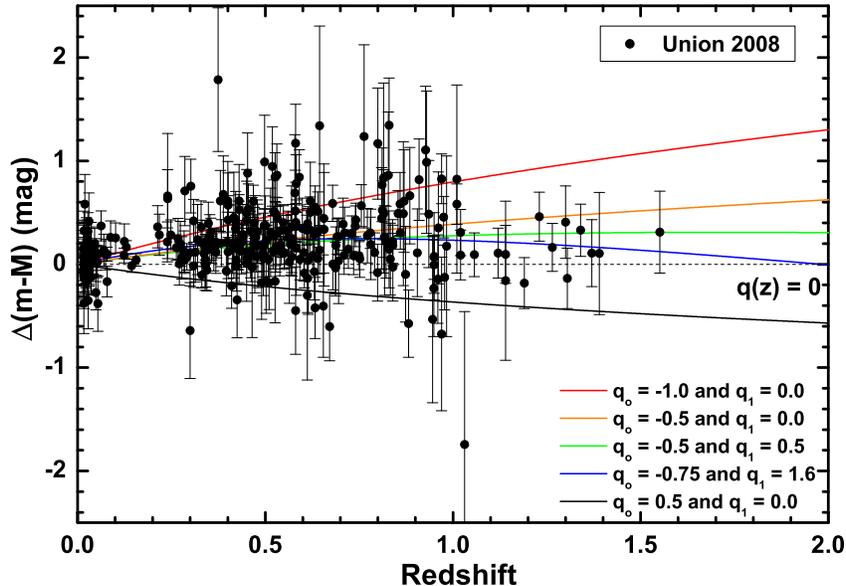}} \caption{Residual
magnitude  versus redshift is shown for 307 SNe type Ia from SCP
Union compilation. Data and kinematic models of the expansion
history are shown relative to an eternally coasting model,
$q(z)\equiv 0$.}
\end{figure}

Although generalizable for non-zero curvature,  Eq. (1) is not a
crude approximation as one may think at first sight.  In the
framework of a flat FRW type universe, it is an exact expression for
the luminosity distance which depends on the epoch-dependent
deceleration parameter, $q(z)$, as well as on the present Hubble
constant, $H_0$.  The simplest way to work with the coupled
definitions (1) and (2) as a kinematic  model for the SN type Ia
data is by adopting a parametric representations for $q(z)$. As one
may check, in the case of a linear two-parameter expansion for
$q(z)=q_0 +z{q_1}$ \cite{Ries04}, the integral (1) can be
represented in terms of a special function as (see \cite{CunLim08})
\begin{eqnarray}\label{eq:dLKin}
D_L(z) &=& \frac{(1+z)}{H_0}e^{q_1}q_1{^{q_0-q_1}}
[\gamma{({q_1-q_0},(z+1){q_1})} \nonumber \\
&& \,\, - \gamma{({q_1-q_0},{q_1})}],
\end{eqnarray}
where ${q_0}=q(z=0)$ is the present value  of the deceleration
parameter, ${q_1}$ is the derivative in the redshift evaluated at
$z=0$, and $\gamma$ is the incomplete gamma function with the
condition ${q_1-q_0}>0$ must be satisfied (for more details see
Cunha and Lima\cite{CunLim08}).

Now, by using the above expressions we may get information about
$q_0$, $q_1$ and, therefore, about the global behavior of $q(z)$.
Note also that a positive transition redshift, $z_t$, is  obtained
only for positive signs of $q_1$ (the variation rate of $q_0$) since
$q_0$ is negative and the dynamic transition (from decelerating to
accelerating) happens at $q(z_t)=0$, or equivalently,
$z_t=-q_0/q_1$.

\begin{figure*}[htb]
\centerline{\epsfysize=45mm\epsffile{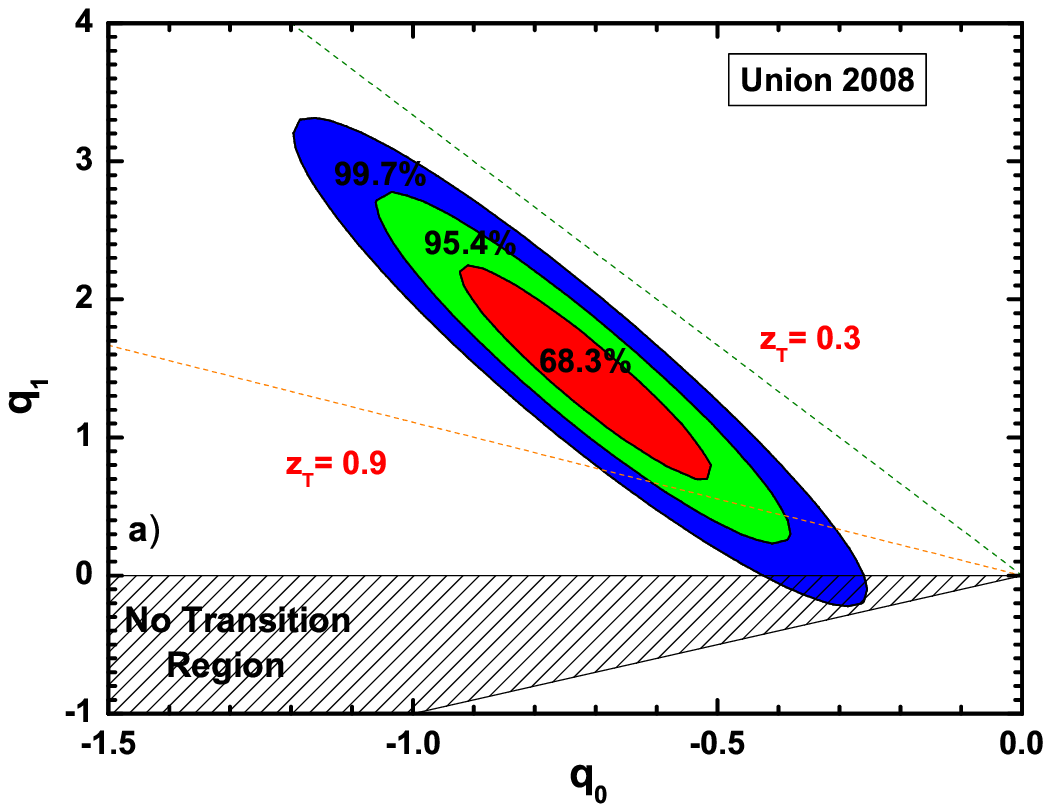}
\epsfysize=45mm\epsffile{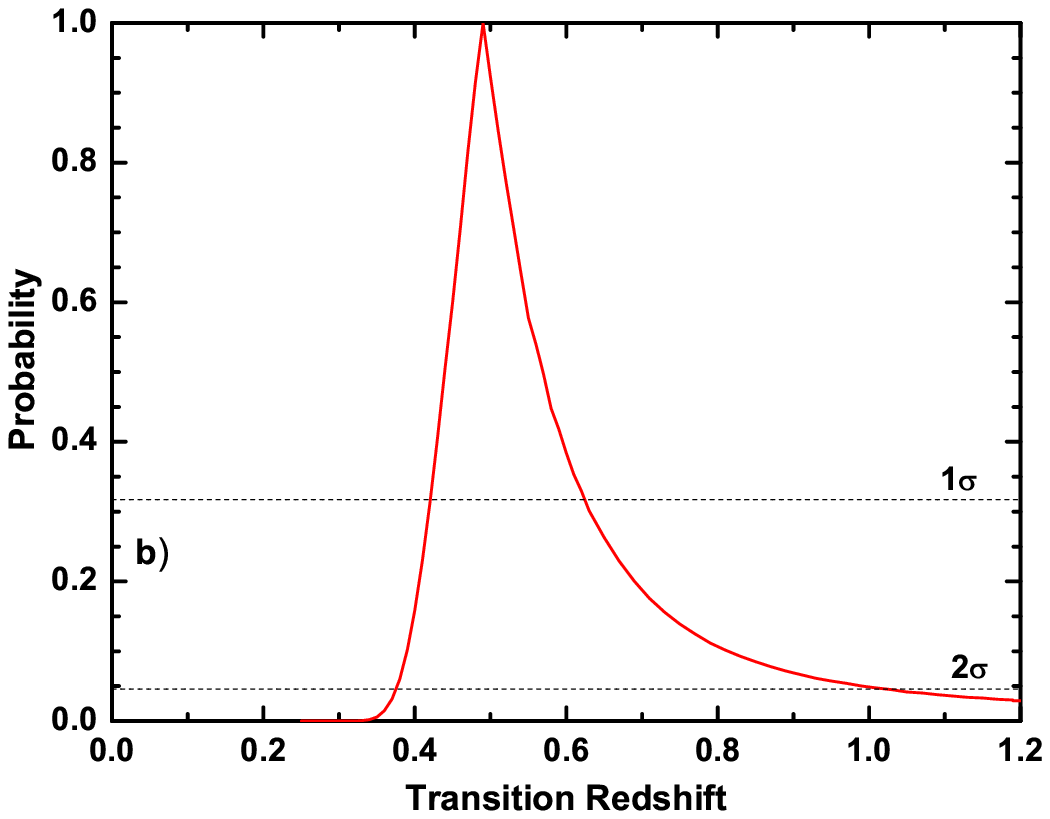}
\epsfysize=45mm\epsffile{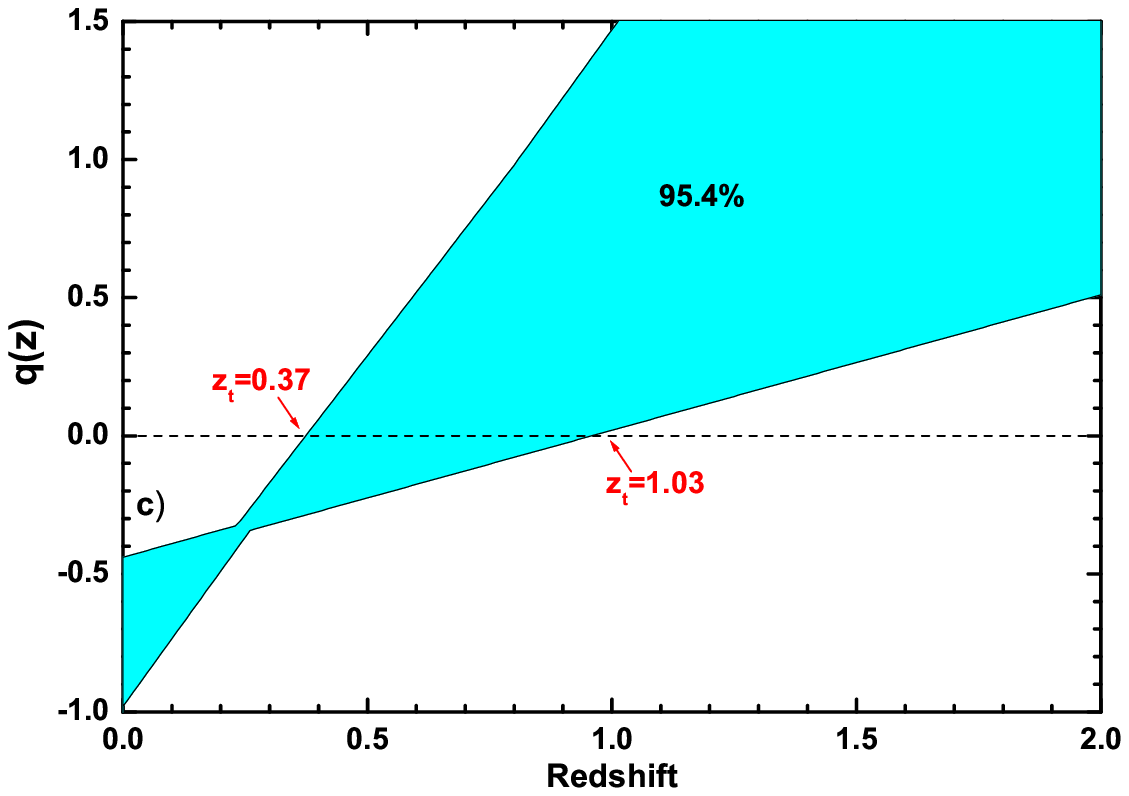}} \caption{{\bf a)} The likelihood
contours in the $q_0 - q_1$ plane for 307 SNe type Ia data. The
contours correspond to 68\%, 95\% and 99\% confidence levels. The
best fit to the pair ($q_0,q_1$) = ($-0.73,1.5)$. {\bf b)}
Probability function for the past transition redshift for a
two-parameter model of the expansion history, $q(z)=q_0+q_1z$. Our
analysis furnishes the best fit $z_t = 0.49^{+0.14}_{-0.07}$
($1\sigma$) $^{+0.54}_{-0.12}$ ($2\sigma$). {\bf c)} Evolution of
the decelerating parameter as a function of the redshift. In the
panel the shadowed region means 2$\sigma$ level for the SCP Union
sample. The dotted horizontal lines represent the coasting model
($q(z)\equiv 0$).}
\end{figure*}

In the statistical analysis below we consider the most complete
dataset we have right now, SCP Union sample\cite{Kowalski08}. The
Union SNe compilation is a new dataset of low-redshift nearby-
Hubble-flow SNe and new analysis procedures to work with several
heterogeneous compilations SNe Ia. It includes 13 independent sets
with SNe from the SCP, High-z Supernovae Search (HZSNS) team,
Supernova Legacy Survey and ESSENCE Survey, the older datasets, as
well as the recently extended dataset of distant supernovae observed
with HST. After selection cuts, the robust compilation obtained is
composed by 307 SNe Ia events distributed over the redshift interval
$0.015 \leq z \leq 1.55$.

\section{Statistical analysis and Results}

The analysis below is based on  the luminosity distance as given by
[1]  with the  ``linear expansion" for $q(z)$. The primary aim  is
to limit the parameters $q_0$ and $q_1$ by using the Union
compilation data as above discussed. All the results are derived by
marginalizing the likelihood function over the nuisance parameter,
$H_0$, thereby obtaining the contours and the associated
probabilities.

In Figure 1 we show the theoretical predictions of the kinematic
approach to the residual Hubble diagram with respect to an eternally
coasting Universe model ($q(z)\equiv0$). The different models are
characterized by the selected values of $q_0$ and $q_1$, as depicted
in the diagram. Let us consider the maximum likelihood that can be
determined from a $\chi^2$ statistics for a given set of parameters
$(H_0, q_0, q_1)$. In what follows we investigate the bounds arising
on the empirical $q(z)$ parameters and the probability of the
redshift transition for each SNe type Ia sample. By marginalizing
the likelihood function over the nuisance parameter, $H_0$, the
contours and the probabilities of the transition redshift for each
sample are readily computed.

In Figure 2a, we see that the SCP Union sample strongly favors a
Universe with recent acceleration ($q_0<0$) and previous
deceleration ($dq/dz>0$). With two free parameter the confidence
region is $0.7 \leq q_1 \leq 2.2$ and $-0.93 \leq q_0 \leq -0.52$
with ($68\%$) confidence level. It should be remarked the presence
of a forbidden region forming a trapezium.  The horizontal line in
the top is defined by $q_1=0$ which leads to an infinite (positive
or negative) transition redshift while the segment at $45^{o}$ is
the infinite future ($z_t = -1$). The values of $z_t$ associated
with the horizontal segment in the bottom are always smaller than -1
(no transition region), in fact $-1.5 \leq z_t \leq -1$. Finally,
one may conclude that the vertical segment is associated with $z_t
\leq -1.5$, thereby demonstrating that the hachured trapezium is
actually a  physically forbidden region.

In Figure 2b (the center panel) one may see the probability of the
associated transition redshift $z_t$, defined as $q(z_t)=0$. It has
been derived by summing the probability density in the $q_0$ versus
the ${dq/dz}$ plane along lines of constant transition redshift,
$z_t = -q_0/(dq/dz)$. The resulting analysis yields $z_t =
0.49^{+0.135}_{-0.07}$ ($1\sigma$) $^{+0.54}_{-0.12}$ ($2\sigma$)
for one free parameter which is in reasonable agreement with the
value $z_t = 0.46 \pm 0.13$ \cite{Ries04}. In our analysis, the
asymmetry in the probability of $z_t$ is produced by a partially
parabolic curve obtained when $\chi^2$ is minimized. For this panel
the central value $z_t=0.49$ does not agree with the cosmic
concordance $\Lambda$CDM model in $68.3$\% confidence level.
However, it agrees with $95.4$\% ($2\sigma$) for this approach. Note
that $z_t= 0.3$ ($z_t$ for flat $\Lambda$CDM with $\Omega_m\simeq
\Omega_{\Lambda}\simeq 0.5$) is outside of the allowed region with
$95\%$, while, $z_t= 0.9$ (flat $\Lambda$CDM with $\Omega_m\simeq
0.2$ and $\Omega_{\Lambda}\simeq 0.8$) are well inside for the Union
sample.

\begin{table}[htbp]
\caption{Limits to the transition redshift $z_t$.}
\label{tables1}%tab2
\begin{center}
\begin{tabular}{@{}cccc@{}}
\hline Sample (data) & $z_t$ (best-fit) & Confidence ($1\sigma$) &
$\chi^2_{min}$
\\ \hline\hline
Gold 2004 (157)&$0.46$& $0.33 \leq z_t\leq 0.59$&$176$ \\
%\\
Astier 2006 (115)&$0.61$&$0.40\leq z_t\leq 4.29$&$113$ \\
%\\
Gold 2007 (182)&$0.43$&$0.38\leq z_t\leq 0.52$&$156$ \\
%\\
Davis 2007 (192) &$0.60$&$0.49\leq z_t\leq 0.88$&$195$ \\
%\\
{\bf{This paper (307)}}&{\boldmath{$0.49$}}&{\boldmath{$0.42\leq z_t\leq 0.63$}}&{\boldmath{$310$}} \\
\hline
\end{tabular}
\end{center}
\end{table}

At this point it is interesting to investigate how our analysis
constrains the redshift evolution of the decelerating parameter
itself. The basic results are displayed in Fig. 2c, we see the
evolution of the decelerating parameter as a function of the
redshift for the parametrization $q(z)=q_0+q_1z$. The shadowed
region denotes the 2$\sigma$ region. The data favor the recent
acceleration ($q_0<0$) and past deceleration ($q_1>0$) with high
confidence level.

It is also interesting to compare the results derived here with
another independent analyses. The first constraints using this
parametrization was obtained by Riess and collaborators from 157 SNe
the transition redshift was constrained to be at $z_t = 0.46\pm
0.13$ \cite{Ries04}. More recently, using 182 SNe Riess {\it{et
al.}} obtained $z_t = 0.43\pm 0.07$ for the probability density in
the $q_0$ vs. $q_1$ plane along lines of constant transition
redshift \cite{Riess07Lowas08}. In an early paper, we studied this
parametrization to three samples. For Astier data set 2006 we
obtained $z_t = 0.61^{+3.68}_{-0.21}$, Gold sample 2007 the
constraints were $z_t = 0.43^{+0.09}_{-0.05}$, and Davies data set
2007 $z_t = 0.60^{+0.28}_{-0.11}$. The asymmetries in our errors
preserve the asymmetries in the $\chi^2$ analysis \cite{CunLim08}.
Besides that, Bayesian analysis was implemented to study this
kinematic scenario by Elgar{\o}y and Multam\"{a}ki
\cite{ElgaMult06}, and, a kinematical study from type Ia supernovae
and X-ray cluster gas mass fraction measurements, by combining them
they obtain significantly tighter results than using the SNe sample
alone \cite{Rapetti07}.

In  Table 1 we summarize the recent results to the transition
redshift $z_t$ in the kinematic approach derived from different
samples of SNe Ia data. As shown there, for phenomenological law
$q(z)= q_0 + q_{1}z$ the limits were derived separately for each
sample of SNe type Ia data.

\section{Conclusions}

In this paper we have discussed  the transition redshift obtained
from a kinematical approach within a flat FRW standard line element.
Our study strongly favors a Universe with recent acceleration
($q_0<0$) and previous deceleration ($dq/dz>0$) for an analysis
which is independent of the matter-energy content of the Universe
based on the phenomenological law, $q(z) = q_0 + q_1z$, and the SCP
data compilation \cite{Kowalski08}. In our analysis we use the
analytical expression to the distance luminosity \cite{CunLim08}, as
well as, the excluded regions with $z_t<-1$ (figure 2a).

The likelihood function for the transition redshift was also
discussed. In this case, the confidence regions in the bidimensional
space parameter ($q_0,q_1$) do not cross the physically forbidden
region. Our analysis  provides an independent evidence for a
dynamical model in which a kinematic transition phase
deceleration/acceleration happened at redshifts smaller than unity.
Hopefully, the constraints on $z_t$ will be considerably improved in
the near future with the increasing of supernova data at
intermediate and high redshifts.

\section*{Acknowledgments}

I am grateful to J. A. S. Lima, J. F. Jesus and V. C. Busti for
helpful discussions. This work was supported by a FAPESP fellowship
No. 05/02809-5.

\end{document}